\documentclass[12pt]{iopart}
\usepackage{graphicx}
\usepackage{bm}
\usepackage{amssymb}
\usepackage{setstack}
\def\implies{\Rightarrow}
\begin{document}
\title[Production and decay of evolving horizons]
{Production and decay of evolving horizons}%

\author{%
Alex B. Nielsen \footnote[2]{alex.nielsen@canterbury.ac.nz}}
\address{Department of Physics and Astronomy,\\
University of Canterbury,\\
Private Bag 4800,
Christchurch, New Zealand}

\author{%
Matt Visser \footnote[1]{matt.visser@mcs.vuw.ac.nz}}
\address{School of Mathematics, Statistics, and Computer Science, \\
Victoria University of Wellington, \\
P.O.Box 600, Wellington, New Zealand}

\def\Nordstrom{Nordstr\"om}
\def\Painleve{Painlev\'e}
\begin{abstract}

  We consider a simple physical model for an evolving horizon that is
  strongly interacting with its environment, exchanging arbitrarily
  large quantities of matter with its environment in the form of both
  infalling material and outgoing Hawking radiation.  We permit fluxes
  of both lightlike and timelike particles to cross the horizon, and
  ask how the horizon grows and shrinks in response to such flows. We
  place a premium on providing a clear and straightforward exposition
  with simple formulae.

  To be able to handle such a highly dynamical situation in a simple
  manner we make one significant physical restriction --- that of
  spherical symmetry --- and two technical mathematical restrictions:
  (1) We choose to slice the spacetime in such a way that the
  space-time foliations (and hence the horizons) are always
  spherically symmetric. (2) Furthermore we adopt
  \Painleve--Gullstrand coordinates (which are well suited to the
  problem because they are nonsingular at the horizon) in order to
  simplify the relevant calculations. Of course physics results are
  ultimately independent of the choice of coordinates, but this
  particular coordinate system yields a clean physical interpretation
  of the relevant physics.

  We find particularly simple forms for surface gravity, and for the
  first and second law of black hole thermodynamics, in this general
  evolving horizon situation. Furthermore we relate our results to
  Hawking's apparent horizon, Ashtekar \emph{et al.}'s isolated and
  dynamical horizons, and Hayward's trapping horizon.  The evolving
  black hole model discussed here will be of interest, both from an
  astrophysical viewpoint in terms of discussing growing black holes,
  and from a purely theoretical viewpoint in discussing black hole
  evaporation via Hawking radiation.

\bigskip

\centerline{gr-qc/0510083; 18 October 2005; 26 October 2005; 
Revised 5 May 2006; \LaTeX-ed \today}

\centerline{Classical and Quantum Gravity 23 (2006) 4637-4658.}

\end{abstract}
\pacs{04.20.-q; 04.20.Cv}
\maketitle
\def\d{{\mathrm{d}}}
\def\~t{\tilde{t}}

\section{Introduction}

Throughout the 1970's the black hole spacetimes considered in the
scientific literature were typically globally stationary and
asymptotically flat --- with true mathematically precise event
horizons [absolute horizons]. When accretion was considered it was
most often handled in the test-matter limit, where the
self-gravitation of the accreting matter was ignored.  After the
accretion rate was calculated in this approximation one might feed
this back into the black hole geometry to allow the spacetime to
evolve in a quasi-static manner, but the horizon itself was almost
always treated as though the idealized Schwarzschild or Kerr event
horizon captured almost all of the relevant physics. The ``Membrane
paradigm'' is perhaps the apex of this quasi-stationary
approach~\cite{Membrane}.  Similarly, calculations concerning Hawking
radiation are typically carried out in the test-field limit, where the
gravitational field generated by the outgoing Hawking flux is
ignored~\cite{Hawking}.  Once the Hawking flux is calculated, this is
set equal to the mass loss rate of the underlying black hole, with the
black hole again evolving in a quasi-stationary manner~\cite{Birrell}.

Those explicit models where exact solutions of the Einstein equations
were used to investigate black hole horizons beyond the
quasi-stationary approximation generally made rather strong
assumptions on the nature of the matter crossing the horizon. For
instance, in Oppenheimer--Snyder collapse~\cite{Oppenheimer} one is
limited to dealing with zero pressure dust, while in the Vaidya
solutions~\cite{Vaidya} one is limited to lightlike radiation crossing
the horizon~\cite{Exact} (either incoming radiation, or outgoing
radiation, but not both).

More recently, considerable work has been done on relaxing these
assumptions in various ways.  Motivated partly by advances in
numerical relativity, Ashtekar \emph{et al.}~\cite{Ashtekar}, and
independently Hayward~\cite{Hayward}, have developed formalisms that
address what they variously call ``isolated'', ``dynamical'', and
``trapping'' horizons. These modifications of the notion of event
horizon exhibit some but not all of the properties of the ``apparent''
horizon defined (for instance) in~\cite{Hawking-Ellis}.  (For more
technical details, see the recent review~\cite{Booth}.) Herein we
shall present a particularly simple framework that, while assuming
spherical symmetry, does not require asymptotic flatness (or more
importantly any notion of stationarity) either of the global spacetime
or of the black hole itself. Similarly, no \emph{a priori} constraints
are placed on the nature or quantity of matter crossing the horizon,
in either direction.  Thus the framework provides a useful
illustration and testing ground for some of the ideas recently
popularized by Ashtekar \emph{et al.} and Hayward.

The choice of \Painleve--Gullstrand coordinates $(\~t,r,\theta,\phi)$
rather than the more familiar Schwarzschild coordinates
$(t,r,\theta,\phi)$ is motivated by a desire to clarify issues such as
how the black hole is able to grow despite the supposed ``frozen''
nature of Schwarzschild time at the horizon, and the related issue of
wanting to calculate quantities at the horizon using a coordinate
system that does not break down at the horizon.  The use of $\~t$
distinguishes the \Painleve--Gullstrand time coordinate from the
Schwarzschild time coordinate $t$. These two coordinates agree at
spatial infinity where they correspond to the proper time of observers
at infinity.  We do \emph{not}, however, require that such an
asymptotically flat region need exist.

We also include some discussion of the relation between our evolving
horizons, apparent horizons, and the isolated/dynamical/trapping
horizons of Ashtekar \emph{et al.} and Hayward.  It has recently been
conjectured that in a fully quantum gravitational universe true
mathematically precise event horizons will not form, only long-lived
apparent horizons~\cite{Dublin}.  This fact has led to claims of a
solution to the black hole information paradox~\cite{Dublin,
  Ashtekar:fuzzy, Hayward:disinformation, Hayward-special}.  While the
present work does not directly shed any light on this issue, it does
provide a classical arena wherein such ideas can be
discussed.\;\footnote{For a related quantum framework phrased in terms
  of the spherically symmetric sub-sector of the coupled
  Einstein-scalar field system see~\cite{Viquar}.}

The organization of the paper is as follows.  In section \ref{S:pg} we
derive the form of the time dependent, spherically symmetric metric in
\Painleve--Gullstrand coordinates and in sections \ref{S:evolve}
and \ref{S:nec} we use this formulation to derive equations for the
rate of change of the horizon radius and area.  In sections
\ref{S:naturaltet} and \ref{S:generaltet} we calculate various fluxes
and stresses at the horizon using a particular choice of tetrad basis.
In sections \ref{S:trapping} and \ref{S:generic} we relate this work
to that done by Ashtekar \emph{et al.}, and Hayward, and discuss the
unified formalism for expanding, stationary, and evaporating horizons.

\section{\Painleve--Gullstrand coordinates}
\label{S:pg}

It is well known that any spherically symmetric gravitational field
can always be written:
\begin{equation}
\d s^2 =
g_{tt}(r,t) \,\d t^2 + 2 g_{rt}(r,t) \,\d r\,\d t 
+ g_{rr}(r,t)\, \d r^2 + R(r,t)^2 \; \d\Omega^2.
\end{equation}
The remaining coordinate freedom in the $r$--$t$ plane can then be
used to reduce the four independent functions above to two. For
instance, it is also well known that a \emph{static}, spherically
symmetric gravitational field can always be written, at least locally,
in terms of curvature coordinates (\emph{aka} Schwarzschild
coordinates), as~\cite{Lorentzian,dbh1}
\begin{equation}
\d s^2 = -e^{-2\Phi(r)}\left[1-{2m(r)\over r}\right] \d t^2
+ {\d r^2\over 1-2m(r)/r} + r^2 \; \d\Omega^2,
\end{equation}
and furthermore for static spacetimes this representation is quite
useful --- and in the absence of horizons is even globally useful.
Then for a \emph{time-dependent} spherically symmetric gravitational
field it is natural to write
\begin{equation}
\label{tdepsch}
\d s^2 = -e^{-2\Phi(r,t)}\left[1-{2m(r,t)\over r}\right] \d t^2
+ {\d r^2\over 1-2m(r,t)/r} + r^2 \; \d\Omega^2.
\end{equation}
Indeed, we know that any general spherically symmetric metric can
be written in terms of two unknown functions and here we are just
making them time dependent. So no one can stop you from adopting
such a coordinate system, it's just that (as we shall soon argue)
this particular coordinate system is less useful than one might at
first suppose.

Now for any given slicing of spherically symmetric spacetime into
hypersurfaces, and in particular for the choice above, one can look
for the existence of apparent horizons in the standard way (see, for
instance,~\cite{Hawking-Ellis}).  That the apparent horizons depend
on the slicing of spacetime was shown most clearly in reference
\cite{Slicing} where non-symmetric slicings of Schwarzschild spacetime
were found for which \emph{no} apparent horizons exist. An explicit
example of this can be found in \cite{Schnetter:2005ea}. Because of
this particular ``feature'' of apparent horizons both Ashtekar
\emph{et al.}, and Hayward, when discussing general spacetimes, have
to introduce considerable technical machinery to avoid pathologies due
to potentially ill-behaved slicing of the spacetime.

In contrast, in this article we will restrict ourselves to spherically
symmetric spacetimes \emph{and} spherically symmetric slicings. So
much, though not all, of the technical superstructure simplifies. (The
whole point of the present article is that we want to retain just
enough technical machinery to do the job, and eliminate as many
complications as possible.) Then for the coordinate choice in equation
\eref{tdepsch} the apparent horizon is informally defined by the
implicit relation $2m(r,t)/r=1$. (We will be more precise and formal
below.)  That is:
\begin{equation}
2 m(r_H(t), t) = r_H(t).
\end{equation}

In the Schwarzschild metric $m(r,t) = M$, where $M$ is a constant, and
this just corresponds to the familiar Schwarzschild radius.  But once
you start asking questions about the evolution $r_H(t)$ of this
apparent horizon you are plagued with multiple divide by zero errors,
as the matrix of metric coefficients, when written in these particular
coordinates, is a singular matrix at the horizon. The ultimate reason
for this behaviour is that Schwarzschild coordinates are reasonably
good (or at least not intolerably bad) for probing the geometry of a
static black hole, but are not particularly good (in fact, downright
awful) for probing the future horizon of an evolving black hole.

To deal with a dynamical black hole, we should use a coordinate system
that is well behaved at the apparent horizon. For this article we will
use \Painleve--Gullstrand coordinates which are particularly simple
and lead to a particularly nice physical picture. (Other non-singular
coordinate choices such as Eddington--Finkelstein, generalised Vaidya,
or double-null coordinates are possible and the physics is entirely
equivalent.~\footnote{For an early article presenting an interesting
  formulation in terms of double null coordinates
  see~\cite{Bergmann}.} We concentrate on \Painleve--Gullstrand
coordinates because they give a particularly nice form to the
equations.)

Starting from (\ref{tdepsch}) we perform the change of coordinates
$t\rightarrow \~t(t, r)$ where $\~t$ will be the
\Painleve--Gullstrand time. Thus
\begin{equation}
\d \tilde t
= \frac{\partial \~t}{\partial t}  \;\d t + \frac{\partial
\~t}{\partial r}\; \d r
\equiv \dot{\~t}\;\d t + \~t '\; \d r.
 \end{equation}
Substituting in for $\d t$,
\begin{equation} \fl 
\d s^{2} =
-e^{-2\Phi(r,t)}\left(1-\frac{2m(r,t)}{r}\right)\;
\left(\frac{1}{\dot{\~t}}\;\d\~t -\frac{\~t'}{\dot{\~t}}\;\d r\right)^{2}
+\frac{dr^{2}}{1-2m(r,t)/r}+r^{2}d\Omega^{2}.
\end{equation}
Expanding out and demanding that $g_{rr}=1$ gives the condition
\begin{equation} \label{tdottdash}
\~t'=\pm
\frac{\sqrt{2m(r,t)/r}}{1-2m(r,t)/r} \; e^{\Phi(r,t)}  \;\; \dot{\~t}.
\end{equation}
That this partial differential equation, which is simply a first-order
linear homogeneous equation for $\tilde t(t,x)$, always has a unique
solution can be shown [for instance] by the method of characteristic
curves~\cite{Courant-Hilbert}\;\footnote{Of course this uniqueness
  holds only subject to the specific choice $g_{rr}=1$, and a suitable
  boundary condition (such as $t'\to t$ when $r\to\infty$).  What is
  certainly not unique is the general process of constructing a
  nonsingular coordinate system at the horizon.  Other coordinate
  systems, such as Eddington--Finkelstein coordinates, are possible
  and will lead to similar results, that are likely however to differ
  in small technical details.}.  Imposing (\ref{tdottdash}) now leads
to
\begin{equation} g_{\~t r} = \pm \frac{e^{-\Phi(r,t)}}{\dot{\~t}}\sqrt{2m(r,t)/r},
\end{equation}
which can be written as
\begin{equation}
g_{\~tr} \equiv \pm c(r, \~t)\sqrt{2m(r,\~t)/r}
\equiv v(r,\~t).
\end{equation}
This implicitly defines $c(r,\~t)$ and leads to
\begin{equation} g_{\~t\~t}=-(c^{2}-v^{2}). \end{equation}
Thus the final metric, now in \Painleve--Gullstrand form, becomes
\begin{equation}
\d s^2 = - c(r,\~t)^2\; \d \~t^2 + [\d r + v(r,\~t)\,\d \~t\,]^2 
+ r^2\; \d\Omega^2,
\end{equation}
or equivalently
\begin{equation} \label{PGmetric}
\d s^2 = - [c(r,\~t)^2-v(r,\~t)^2]\d \~t^2 + 2 v(r,\~t) \; \d r\; \d \~t 
+ \d r^2 + r^2 \; \d\Omega^2.
\end{equation}
That is, any spherically symmetric space-time, regardless of whether
it is static or not, can always locally be put into this form.  Note
that, as is usual in \Painleve--Gullstrand coordinates, surfaces of
constant $\~t$ are spatially flat, and there is a natural notion of
``outwards'' and ``inwards'' associated with increasing or decreasing
the $r$ co-ordinate.  Furthermore, four-dimensional asymptotic
flatness would correspond to imposing $c \rightarrow 1$ and $v
\rightarrow 0$ as $r \rightarrow \infty$. (We will \emph{not} need to
impose asymptotic flatness in the discussion that follows.) Note also
that the static Schwarzschild case would be given by setting $c=1$
everywhere, while $v(r)=\sqrt{2M/r}$, with $M$ a constant.  Explicitly
we have
\begin{equation}
g_{ab} = \left[ \begin{array} {c|c}-[c^2-v^2]&v_j\\ 
\hline v_i& h_{ij}\end{array} \right],
\end{equation}
and
\begin{equation}
g^{ab} = \left[ \begin{array} {c|c}-c^{-2}&v^j/c^2\\ 
\hline v^i/c^2& h^{ij}-v^iv^j/c^2\end{array} \right],
\end{equation}
where $h_{ij}$ is the metric of flat Euclidean 3-space in spherical
polar coordinates.  The ingoing and outgoing radial null curves are
defined by $\d s^2=0$ and equation (\ref{PGmetric}) gives
\begin{equation}
{\d r\over\d \~t} = - v(r,\~t) \pm c(r,\~t).
\end{equation}
Thus we can define the location of what we will call the evolving
horizon by the very simple and intuitive condition
\begin{equation}
c(r,\~t) = v(r,\~t),
\end{equation}
which has the simple physical interpretation that when $c(r,\~t) <
v(r,\~t)$ the outgoing light ray is being dragged backwards to smaller
values of $r$. This implicitly defines a function $r_H(\~t)$ such that
\begin{equation}
v(r_H(\~t),\~t) = c(r_H(\~t),\~t),
\end{equation}
which is equivalent to
\begin{equation}
2m(r_H(\~t),\~t) = r_H(\~t).
\end{equation}
Now the question is; what is the evolution of this function
$r_H(\tilde t)$ in terms of the stress-energy at $r_H(\~t)$?

\section{Surface gravity and the First law}
\label{S:evolve}

Using {\sf Maple} (or some equivalent) to compute the Riemann tensor
of the metric (\ref{PGmetric}) in orthonormal components we calculate
\begin{equation}
R_{\hat\theta\hat\phi\hat\theta\hat\phi} =  {2 m(r,\~t)\over r^3},
\end{equation}
so that $m(r,\~t)$ can be physically identified as the
Hernandez--Misner mass function~\cite{XXX}. Similarly, since we are
dealing with a spherically symmetric spacetime, we could compute the
Hawking--Israel quasi-local mass function
\begin{equation}
m_\mathrm{HI}(r,t) =  {r\over2} 
\left[ 1 - g^{ab} \; \nabla_a r \; \nabla_b r \right],
\end{equation}
which again leads to the mathematical quantity $m(r,t)$ appearing
in the metric being physically identified as the ``mass inside
radius $r$ at time $t$''.

Now since $2m(r,\~t)=r$ at the evolving horizon, we have
 \begin{equation}
 2\dot m(r_H(\~t),\~t) + 2 m'(r_H(\~t),\~t) \dot r_H(\~t) = \dot r_H(\~t),
 \end{equation}
 which we can recast as
 \begin{equation}
 \label{E:mdot}
 \dot m(r_H(\~t),\~t) = {[1- 2 m'(r_H(\~t),\~t)]\over 2}\;  \dot r_H(\~t).
 \end{equation}
 Introducing $\mathcal{A}_H = 4\pi r_H^2$, the area of the evolving
 horizon, we have
\begin{equation}
 \dot m(r_H(\~t),\~t)
 = {1\over 8\pi} {[1- 2 m'(r_H(\~t),\~t)] \over 2 r_H(\~t)} \dot
 \mathcal{A}_H(\~t),
 \end{equation}
 where we have seen that the function $m(r)$ can be physically
 interpreted, due to the spherical symmetry, as the mass contained
 within a radius $r$.  Therefore this has exactly the \emph{form} of
 the first law of black hole mechanics, $dm = {1\over8\pi} \kappa \;
 d\cal{A}$, though now in a completely dynamical context, provided we
 agree to focus on partial derivatives with respect to $\tilde t$
 \footnote{If we were to compute the total time derivative of
   $m(r_H(\tilde t),\tilde t)$, we would obtain
\[
{\d m(r_H(\tilde t),\tilde t)\over \d \tilde t} =  
\dot m(r_H(\tilde t),\tilde t) 
+  m'(r_H(\tilde t),\tilde t) \; \dot r_H(\tilde t) =
{ \dot r_H(\tilde t) \over 2}.
\]
This result, though simple, does not appear to be particularly useful,
and does not seem to have a straightforward interpretation in terms of
black hole thermodynamics. The best one can apparently do in this
regard is to rearrange the above as
\[
{\d m(r_H(\tilde t),\tilde t)\over \d \tilde t} =   
{\dot m(r_H(\tilde t),\tilde t) \over
1 - 2  m'(r_H(\tilde t),\tilde t)}.
\]
},  and also to identify
 \begin{equation}
 \kappa_H(\~t)
 =  {[1- 2 m'(r_H(\~t),\~t)] \over 2 r_H(\~t)}
 \end{equation}
 as the ``surface gravity''.\;\footnote{For a discussion of situations
   wherein the ``entropy= area/4'' law might fail. see for
   instance~\cite{dbh2,dbh3}.}  To justify this interpretation of
 $\kappa_H$, it is useful to first define the outward radial null
 vector
\begin{equation}
\ell^a={\left(\vphantom{\Big{|}}1,c(r,\~t)-v(r,\~t),0,0\right)\over c(r,\~t)},
\end{equation}
and verify that $g_{ab} \;\ell^a\;\ell^b = 0$.  The overall
normalization is chosen to make things simpler below.  It is also
useful to define the inward radial null vector
\begin{equation}
n^a={\left(\vphantom{\Big{|}}1,-c(r,\~t)-v(r,\~t),0,0\right)\over c(r,\~t)},
\end{equation}
and verify that this is also null $g_{ab} \;n^a\;n^b = 0$, and that
$g_{ab} \;\ell^a\;n^b = -2$. (This choice of normalization is the most
``symmetric'' we have been able to find.)

Because of spherical symmetry we must have
\begin{equation}
\ell^b \; \nabla_b \ell^a = \kappa_\ell \; \ell^a; \qquad
n^b \; \nabla_b n^a = \kappa_n \; n^a;
\end{equation}
where the scalars $\kappa_l$ and $\kappa_n$ are defined everywhere on
the spacetime, not just on the evolving horizon.  Computing
$\kappa_\ell$ and $\kappa_n$ for the given $\ell^a$ and $n^a$ yields
\begin{eqnarray}
\kappa_\ell(r,\~t) &= {c'(r,\~t)-v'(r,\~t)\over c(r,\~t)}
\\
&= {1\over2r}{\left[{2m(r,\~t)/ r} - 2 m'(r,\~t)\right]
\over\sqrt{2m(r,\~t)/r}} +
{c'(r,\~t)\over c(r,\~t)}\left[1-\sqrt{2m(r,\~t)
\over r}\right];
\end{eqnarray}
\begin{eqnarray}
\kappa_n(r,\~t) &= {-c'(r,\~t)-v'(r,\~t)\over c(r,\~t)}
\\
&= {1\over2r}{\left[{2m(r,\~t)/ r} - 2 m'(r,\~t)\right] 
\over\sqrt{2m(r,\~t)/r}}+
{c'(r,\~t)\over c(r,\~t)}\left[1+\sqrt{2m(r,\~t)\over r}\right].
\end{eqnarray}
On the evolving horizon $\kappa_\ell$ reduces to
\begin{equation}
\kappa_H(\~t) = {1-2m'(r_H(\~t),\~t)\over2r_H(\~t)},
\end{equation}
a formula which has all the appropriate limits and is compatible up to
normalization with the results in reference~\cite{dbh1}.  In
particular, if the geometry is static (time-independent) then it is
obvious that evolving, apparent, event, isolated, dynamic, and
trapping horizons coincide, and that $\kappa_H$ as defined above
reduces to the standard definition. (In reference~\cite{dbh1} the
spacetime was in addition assumed asymptotically flat, which we do not
need to assume in the current analysis.  The asymptotic flatness was
then used to motivate a particular non-local normalization for surface
gravity. The current definition of ``surface gravity'' is in contrast
local, and only uses information that can be extracted from the
vicinity of the evolving horizon itself.)

Now we want to be dealing with the outermost evolving horizon, which
for simplicity we define in the purely intuitive sense.  Thus for
purely kinematic reasons
\begin{equation}
 {2m(r,\~t)\over r} < 1 \qquad \mathrm{for} \qquad r >
  r_H(\~t),
\end{equation}
which implies
\begin{equation} \label{extremeBH}
\left.\left({2m\over r}\right)'\right|_H \leq 0 
\qquad \implies 
\qquad 
1 - 2 m'(r_H(\~t),\~t) \geq 0.
\end{equation}
Note that in the case of a spacelike evolving horizon the horizon might,
in principle, intersect the constant time slices many times. In such a
situation this condition will only identify the outermost
intersection. See the appendix on the Vaidya spacetime for more
details on this point. Apart from the exceptional case
$1-2m'(r_H(\~t),\~t)=0$, we see that the surface gravity on the
outermost horizon is guaranteed to be positive. Indeed, from the above
we can see that the case where $1-2m'(r_H(\~t),\~t)=0$ corresponds to
$\kappa_H(\~t)=0$, the condition for an extremal horizon.\footnote{For
  example, in the case of the Reissner--{\Nordstrom} spacetime we
  have $m(r) = M-Q^{2}/2r$ and a little bit of algebra is enough to
  show that the condition $1-2m'(r_H)=0$ is equivalent to the
  condition $M^{2}=Q^{2}$, the usual condition for an extremal
  Reissner--{\Nordstrom} black hole. However in more general
  situations the vanishing of the surface gravity is typically taken
  to be the \emph{primary} definition of what one means by
  extremality.}  For a discussion of the subtleties that can occur
once one has multiple nested dynamical horizons, see~\cite{Ivan}.

\paragraph{Aside:} We could also calculate $\kappa_n$, corresponding
to the ingoing null geodesics, on the evolving horizon. Though there
is no difficulty in doing so, there is no clear physical
interpretation of the resulting quantity:
\begin{eqnarray}
\kappa_n(r_H(\~t),\~t)
&=  \kappa_H(\~t)+
{2 c'(r_H(\~t),\~t)\over c(r_H(\~t),\~t)}.
\end{eqnarray}

\paragraph{Aside:} From the defining relations for $\kappa_\ell$ and
$\kappa_n$, the spherical symmetry of the spacetime, and the
normalization relation between $\ell$ and $n$, it is easy to see that
\begin{equation}
n^b \; \nabla_b \ell^a = -\kappa_n \; \ell^a; \qquad
\ell^b \; \nabla_b n^a = -\kappa_\ell \; n^a;
\end{equation}
or equivalently
\begin{equation}
\kappa_n = {n^a \; n^b \; \nabla_b \ell_a \over2}; \qquad
\kappa_\ell = {\ell^a \; \ell^b \; \nabla_b n_a \over 2}.
\end{equation}

\paragraph{Aside:} From the definition of $\kappa_\ell$
\begin{equation}
\ell^b \; \nabla_b \ell^a = \kappa_\ell \; \ell^a
\end{equation}
we see that rescaling the null vector $\ell\to \alpha \ell$ (which
implies $n\to \alpha^{-1} \, n$), will result in
\begin{equation}
\kappa_{(\alpha\,\ell)} = \alpha \;\kappa_\ell + \ell\cdot\nabla \alpha.
\end{equation}
So at the evolving horizon, where $\ell\cdot\nabla \to c_H^{-1}\;
\partial_t$, we have
\begin{equation}
\kappa_H(\alpha\;\ell) = \alpha \;\kappa_H(\ell) + {\dot\alpha \over c_H}.
\end{equation}
That is, among all possible normalizations for $\ell$, (and so
implicitly for $n$), the one we have chosen is seen to not only be
``symmetric'' but additionally to eliminate any explicit time
derivatives at the evolving horizon.

\section{Null energy condition}
\label{S:nec}

Now calculate the quantity $G_{ab}\; \ell^a\; \ell^b$. Using {\sf
  Maple} (or some equivalent) it is easy to see that
\begin{equation}
\fl
G_{ab}\; \ell^a\; \ell^b = 
{2\over c(r,\~t) \; r^2} {\dot m(r,\~t)\over\sqrt{2m(r,\~t)/r}} +
 {2\over c(r,\~t)\;r}  c'(r,\~t) 
\left( 1 - \sqrt{2m(r,\~t)\over r} \right)^2,
\end{equation}
which we can rearrange to yield
\begin{equation} \fl
\dot m(r,\~t) = {1\over2} {c(r,\~t)\; r^2} \sqrt{2m(r,\~t)\over r} 
                \; G_{ab}\; \ell^a\;\ell^b
 - c'(r,\~t) \sqrt{2m(r,\~t)\,r} 
  \left( 1 - \sqrt{2m(r,\~t)\over r} \right)^2.
\end{equation}
This mass formula applies at \emph{any} value of $r$. In particular at
the evolving horizon we have the very simple result
\begin{equation}
\dot m(r_H(\~t),\~t) = {1\over2} {r_H^2(\~t)\; c(r_H(\~t),\~t)} 
\; G_{ab}\; \ell^a\;\ell^b.
\end{equation}
Invoking the Einstein equations $G_{ab}=8\pi \; T_{ab}$ (with
$G_N\to1$), we see
\begin{equation}
\dot m(r_H(\~t),\~t) =  {4\pi r_H^2(\~t)\; c(r_H(\~t),\~t)} 
\; T_{ab}\; \ell^a\;\ell^b.
\end{equation}
Note that this only includes the mass change due to flux across the
instantaneous location of the evolving horizon, this does not yet
include the effect due to the motion of the evolving horizon.

But in view of equation \eref{E:mdot} we have
\begin{equation}
 \dot r_H(\~t) = {2\dot m(r_H(\~t),\~t)\over 1 - 2 m'(r_H(\~t),\~t)}
 =  {8\pi r_H^2(\~t)\; c(r_H(\~t),\~t) \; T_{ab}\; \ell^a\;\ell^b 
    \over1-2m'(r_H(\~t),\~t)}.
\end{equation}
Furthermore since we have already seen that $1-2m'(r_H(\~t),\tilde t)
> 0$ for any non-extremal outermost horizon, we deduce that the sign
of $\dot r_H$ is the same as the sign of $T_{ab}\;\ell^a\;\ell^b$.
But this is exactly the combination that enters into the Null Energy
Condition ($T_{ab}\; \ell^a\;\ell^b \ge 0$; the NEC) evaluated on the
apparent horizon.

That is, as long as the NEC is satisfied the horizon cannot shrink.
This is compatible of course with the standard analysis in terms of
the Raychaudhuri equation~\cite{Hawking-Ellis}, but here we can see
the result dropping directly out of one component of the Einstein
equations applied to the evolving horizon.\;\footnote{Similar
  violations of the NEC occur at the throats of dynamically evolving
  wormholes~\cite{evolving-wormholes} --- and one also encounters
  problems similar to those for evolving horizons if one tries to
  describe evolving wormholes in Schwarzschild curvature coordinates.
  Schwarzschild curvature coordinates are ill-adapted to discussing
  the evolution of a wormhole throat.} We emphasise that the NEC,
because it is the weakest of the standard energy conditions, leads to
the strongest form of the singularity theorem~\cite{Roman1,Roman2}.
Because the singularity and area increase theorems can both be phrased
in terms of the NEC we do not really need to consider the WEC, SEC, or
DEC.

\paragraph{Aside:} We can also define the total quasi-local mass
inside the evolving horizon by
\begin{equation}
m_H(\~t) = m(r_H(\~t),\~t) = {r_H(\~t)/2},
\end{equation}
and then
\begin{equation}
{\d m_H\over\d \~t}(\~t) = {\dot m(r_H(\~t),\~t)\over 1 - 2 m'(r_H(\~t),\~t)}
=  {4\pi r_H^2(\~t)\; c(r_H(\~t),\~t) \; T_{ab}\; \ell^a\;\ell^b 
\over1-2m'(r_H(\~t),\~t)}.
\end{equation}
This evolution law now characterizes the total mass change, now with
contributions both from mass flux across the evolving horizon and from
the motion of the evolving horizon itself.

\paragraph{Aside:} In contrast, consider the ingoing radial null
direction. For $G_{ab}\;n^a\;n^b$ we have the result
\begin{equation}
\fl
G_{ab}\; n^a\; n^b = {2\over c(r,\~t) \; r^2} 
{\dot m(r,\~t)\over\sqrt{2m(r,\~t)/r}} +
 {2\over c(r,\~t)\;r}  c'(r,\~t) \left( 1 + \sqrt{2m(r,\~t)\over r} \right)^2,
\end{equation}
which we can rearrange to yield
\begin{equation} \fl
 \dot m(r,\~t) = {1\over2} {c(r,\~t)\; r^2} \sqrt{2m(r,\~t)\over r} 
\; G_{ab}\; n^a\;n^b
 - c'(r,\~t) \sqrt{2m(r,\~t)\,r} \left( 1 + \sqrt{2m(r,\~t)\over r} \right)^2,
\end{equation}
but this is nowhere nearly as useful since its form on the evolving
horizon is not as nice.

\section{Apparent, Isolated, Dynamical, Trapping, and Evolving  Horizons}
\label{S:trapping}

While we have used \Painleve--Gullstrand coordinates above to get
some very simple and explicit results, we note that similar but
coordinate independent definitions of various types of horizon are
given by~\cite{Ashtekar, Hayward, Hawking-Ellis}. To see the relation
between the evolving horizons at $2m(r) = r$ and the
apparent/isolated/dynamical/trapping horizons of
Hawking/Ashtekar/Hayward we need to calculate the expansions
$\theta_{\ell}$ and $\theta_{n}$ of the radial null vectors $\ell^{a}$
and $n^{a}$.  Defining the expansion of the outgoing and ingoing
radial null curves as
\begin{equation}
\theta_{\ell} =
\left[g^{ab} + {n^{a}\ell^{b}+ \ell^{a}n^{b}\over2} \right]
\nabla_{a}\ell_{b}
= \nabla_a \ell^a - \kappa_\ell;
\end{equation}
\begin{equation}
\theta_{n} =
\left[g^{ab} + {n^{a}\ell^{b}+ \ell^{a}n^{b}\over2} \right]\nabla_{a}n_{b}
= \nabla_a n^a - \kappa_n;
\end{equation}
we get
\begin{equation} \theta_{\ell} = \frac{2(c-v)}{rc}
= {2\over r} \left\{1-\sqrt{2m\over r}\right\};
\end{equation}
\begin{equation} \theta_{n} = \frac{-(c+v)}{rc}
=  -{2\over r} \left\{1+\sqrt{2m\over r}\right\}.
\end{equation}
So we see that $\theta_\ell$ changes sign exactly at the evolving
horizon $r_H$. This is enough to see that the evolving horizons of
this article coincide with the standard definition of apparent horizon
--- indeed all you really need for this to hold is to have a
spherically symmetric spacetime with a time slicing that respects the
spherical symmetry.

\bigskip
\noindent
Now  define an  ``Ashtekar horizon'' $H$ as follows:
\begin{enumerate}
\item[\it{i}.]
\emph{H} is a three-dimensional, timelike/ null/ spacelike hypersurface.
\item[\it{ii}.] 
On \emph{H} the expansion of $n^{a}$ is negative $\theta_{n}<0$.
\item[\it{iii}.] 
On \emph{H} the expansion of $\ell^{a}$ is zero $\theta_{\ell}=0$.
\end{enumerate}
The null case represents an ``isolated horizon'' and the spacelike
case represents a dynamical horizon. (This is the astrophysically
relevant case where accretion dominates over Hawking evaporation.)
The timelike case corresponds to a situation where Hawking evaporation
dominates over accretion --- this might be referred to as an
``evaporating horizon''.  From the explicit formulae for the
expansions $\theta_\ell$ and $\theta_n$ it is clear that our evolving
horizon is also a horizon in the sense of Ashtekar \emph{et al.}

Finally, a ``Hayward horizon'', or more precisely a future outer
trapping horizon, replaces condition $i$ with
\begin{equation}
{\cal{L}}_{n}\theta_{\ell}<0.
\end{equation}
But explicitly
\begin{equation}
{\cal{L}}_{n}\theta_{\ell} =
(n\cdot \nabla) \left[  {2\over r} \left\{1-\sqrt{2m\over r}\right\} \right] ,
\end{equation}
and so on the evolving horizon
\begin{equation}
\left.{\cal{L}}_{n}\theta_{\ell}\right|_H =
- {2\over r_H} (n\cdot \nabla) \left[ \sqrt{2m\over r} \; \right]_H=
- {1\over r_H} (n\cdot \nabla) \left[{2m\over r} \; \right]_H
.
\end{equation}
On the evolving horizon we also have
\begin{equation}
 (n\cdot \nabla) \to ( c_H^{-1} \partial_t - 2 \partial_r ),
\end{equation}
whence
\begin{equation}
\left.{\cal{L}}_{n}\theta_{\ell}\right|_H =
- {1\over r_H} 
\left[ {2 \dot m\over cr} -2 \left({2m\over r}\right)' \; \right]_H.
\end{equation}
That is
\begin{equation}
\left.{\cal{L}}_{n}\theta_{\ell}\right|_H =
- {1\over r_H^2} \left[ {2 \dot m\over c} +2 (1-2m') \; \right]_H.
\end{equation}
But we have already seen, on purely kinematic grounds, how to relate
$\dot m|_H$ to $\dot r_H$, so
\begin{equation}
\left.{\cal{L}}_{n}\theta_{\ell}\right|_H =
- {1-2m'_H\over r_H^2} \left[ 2 +{\dot r_H\over c_H} \; \right].
\end{equation}
We have furthermore already argued that $1-2m'|_H>0$, so under the
rather mild condition that $\dot r_H >-2 c_H$ we have
$\left.{\cal{L}}_{n}\theta_{\ell}\right|_H < 0$, in which case our
evolving horizon is also a Hayward-style future outer trapping
horizon.

In the situation where $\dot r_H <-2 c_H$ we have a somewhat unusual
type of spacelike evolving horizon --- in the present context this
would correspond to rapid ``evaporation'' of the horizon. Since the
evolution of the horizon now lies outside the lightcone, evolving
forward along curves of $n^{a}$ does not take us from a region of
positive $\theta_{\ell}$ to negative $\theta_{\ell}$.  Timelike
observers are still instantaneously forced to move inwards, but they
immediately escape the horizon --- so that subsequently some of them
may turn around and escape to ``infinity''.  Indeed, once the observer
exits they will not be able to re-enter the horizon unless the rate of
evaporation slows down.

In short, the very simple and intuitively clear definition of
evolving horizon that we advocate in this article is compatible
with all the standard definitions (apparent, isolated, dynamic,
evaporating, trapping) common in the literature. The one thing the
evolving horizon is \emph{not}, is that it is definitely not an
\emph{event} horizon [absolute horizon]. And
one very important message to take from this entire discussion is
that event horizons [in the strict technical sense] are not
essential to investigating black hole physics~\cite{Dublin,
Essential}.

\section{Energy momentum in a natural tetrad basis}
\label{S:naturaltet}

To understand the energy-momentum fluxes and stresses at the evolving
horizon it is extremely useful to adopt a suitable orthonormal tetrad
basis. Once you have picked a tetrad, which is effectively a choice of
privileged observer, you can begin to ask questions about densities,
fluxes, and stresses measured by that observer.

To construct an appropriate ``natural'' tetrad, start with the
observation that the radial null curves are given by
\begin{equation}
{\d r\over\d t} = - v \pm c,
\end{equation}
which implies that the vector $(1,-v,0,0)^a$ is certainly timelike. In fact
\begin{equation}
V^a = {(1,-v,0,0)\over c} = {\ell^a + n^a\over 2}
\end{equation}
is easily seen to be a timelike unit vector. (In an analogue
interpretation of the metric, $V$ would be the 4-velocity of the
``medium'', and our natural observers would be co-moving with the
medium~\cite{analogue}.)  Similarly,
\begin{equation}
S^a = (0,1,0,0); \qquad S_a = (v,1,0,0);
\end{equation}
is easily seen to be a spacelike unit vector that is orthogonal to
$V$. We take these as the first two elements of the tetrad.
Specifically we set
\begin{equation}
e_{\hat a}{}^a = \left( V^a, S^a, \hat\theta^a, \hat\phi^a \right).
\end{equation}
Note that
\begin{equation}
V^a + S^a = {(1,c-v,0,0)\over c} = \ell^a;
\end{equation}
\begin{equation}
V^a - S^a = {(1,-c-v,0,0)\over c} = n^a,
\end{equation}
where $\ell$ is the outgoing null vector we previously chose to make
the surface gravity computation simple. One may easily compute
\begin{equation}
(V\cdot\nabla) V = {c'\over c} \;S; \qquad 
(V\cdot\nabla) S = {c'\over c} \;V;
\end{equation}
\begin{equation}
(S\cdot\nabla) V = -{v'\over c} \;V; \qquad 
(S\cdot\nabla S) = -{v'\over c} \;V.
\end{equation}
Thus an observer who ``moves with the medium'' is not in geodesic
motion except in the special case $c'=0$.

Writing $\hat t^a = V^{a}$ and $\hat r^a = S^{a}$ and using this
natural tetrad (and its inverse as determined by {\sf
  Maple}~\footnote{We particularly wish to warn readers against
  over-enthusiastic use of the {\sf frame} command in {\sf Maple}, or
  its equivalent in other packages. The {\sf frame} function will
  provide \emph{some} orthonormal tetrad for the specified metric, but
  the orthonormal tetrad it provides is certainly not unique, and may
  not always be the most useful.  In particular, if the spacetime
  contains any form of horizon, the vectors in the orthonormal tetrad
  provided by the {\sf frame} command quite often have divergent
  components at the horizon. The fact that the tetrad is not unique is
  of course an unavoidable consequence of local Lorentz invariance,
  and the divergence of the components of the tetrad is a side effect
  of attempting a ``$v=c$'' Lorentz transformation at the horizon, so
  similar problems will show up in any symbolic computation package.
  We have found it best to determine a suitable well-behaved tetrad
  ``by hand'', and then explicitly feed it to the symbolic
  manipulation program for further computations. }), the orthonormal
components of the Einstein tensor (at any point in the spacetime) are
easily calculated to be
\begin{equation}
G_{\hat t\hat t} = {2m'\over r^2} + {2c'\over c r} {2m\over r};
\end{equation}
\begin{equation}
G_{\hat t\hat r} = -{2c'\over c r } \sqrt{2m\over r};
\end{equation}
\begin{equation}
G_{\hat r\hat r} = -{2m'\over r^2} + {2c'\over c r} 
+ {2 \dot m\over c r^2} {1\over\sqrt{2m/r}}.
\end{equation}
The only ``complicated'' component of the Einstein tensor is the
transverse one:
\begin{eqnarray}
\fl
G_{\hat\theta\hat\theta} = G_{\hat\phi\hat\phi} &=&
-{m''\over r} +{c''\over c} \left( 1 - {2m\over r}\right) 
- {3c' m'\over cr} + {c'\over cr}\left(1+{m\over r} \right)
\\
\fl
&& +\sqrt{2m\over r} \left\{
{\dot m\over4mcr} \left[1+ {2c'r\over c} -{rm'\over m} \right]
+
{1\over2c} \left[ {\dot m'\over m} + {2\dot c'\over c} 
- {2\dot c c'\over c^2}\right]
\right\}.
\nonumber
\end{eqnarray}
The Einstein tensor at the evolving horizon simplifies rather drastically
\begin{equation}
G_{\hat t\hat t}|_H = {2m'\over r^2} + {2c'\over c r};
\end{equation}
\begin{equation}
G_{\hat t\hat r}|_H = -{2c'\over c r };
\end{equation}
\begin{equation}
G_{\hat r\hat r}|_H = -{2m'\over r^2} + {2c'\over c r} 
+ {2 \dot m\over c r^2};
\end{equation}
with the only ``complicated'' component being
\begin{eqnarray}
G_{\hat\theta\hat\theta}|_H = G_{\hat\phi\hat\phi}|_H &=&
-{m''\over r}  - {3c' m'\over cr} + {3c'\over 2cr}\
+ {\dot m\over 2cr^2} \left[ 1+ {2c'r\over c}- 2m'\right]
\nonumber
\\
&& +{1\over c} \left[{ \dot m'\over r} + {\dot c'\over c} 
- {\dot c c'\over c^2}\right].
\end{eqnarray}
Now write the Einstein equations as $G_{ab} = 8\pi G_N \; T_{ab}$, and
adopt units where $G_N=1$. Then (at the evolving horizon)
\begin{equation}
\rho_H = T_{\hat t\hat t} = {G_{\hat t\hat t}\over 8\pi} 
= {m'\over4\pi r^2} +{c'\over4\pi r c};
\end{equation}
\begin{equation}
f_H = T_{\hat t\hat r} = {G_{\hat t\hat r}\over 8\pi} 
= -{c'\over4\pi r c};
\end{equation}
\begin{equation}
p_{r,H} = T_{\hat r\hat r} = {G_{\hat r\hat r}\over 8\pi} =
 -{m'\over 4\pi r^2} + {c'\over 4\pi r c } + {\dot m\over 4\pi r^2 c} ;
 \end{equation}
 with again the only real complication being
 \begin{eqnarray}
 \fl
p_{t,H} = T_{\hat\theta\hat\theta}= T_{\hat\phi\hat\phi} 
= {G_{\hat\theta\hat\theta}\over 8\pi}
 &=&
-{m''\over 8\pi\;r}  - {3c' m'\over 8 \pi \; cr} + {3c'\over 16\pi\;cr}\
\\
\fl
&&
+ {\dot m\over16\pi\;cr^2} \left[ 1+ {2c'r\over c}-2 m'\right]
+{1\over8\pi\;c} \left[ {\dot m'\over r} + {\dot c'\over c} 
- {\dot c c'\over c^2}\right].
 \nonumber
\end{eqnarray}
Note that these must be interpreted as the density, flux, and
pressures \emph{as measured by the particular observer moving with
  4-velocity $V^a$}. In terms of these quantities we have the simple
results
 \begin{equation}
 \dot m = 4\pi r^2 c \; (\rho_H+p_{r,H}+2f_H) 
= 4\pi r^2 c \; [T_{ab} \ell^a \ell^b]_H,
 \end{equation}
 and
 \begin{equation}
 m' = 4\pi r^2 c \; (\rho_H+f_H) = 4\pi r^2 c \; [T_{ab} \ell^a V^b]_H.
 \end{equation}
This now implies that the evolving horizon
shifts according to the formula
\begin{equation}
\dot r_H = 
{8\pi c_H r_H^2 (\rho_H+p_{r,H}+2f_H)\over 1-8\pi r_H^2(\rho_H+f_H)}
= 
{8\pi c_H r_H^2 \; [T_{ab}\ell^a\ell^b]_H 
\over1-8\pi r_H^2 [T_{ab} V^a \ell^b]_H};
\end{equation}
and similarly
\begin{equation}
\dot m_H = {4\pi c_H r_H^2 (\rho_H+p_{r,H}+2f_H)
\over 1-8\pi r_H^2(\rho_H+f_H)}
= {4\pi c_H r_H^2 \; [T_{ab}\ell^a\ell^b]_H
\over1-8\pi r_H^2 [T_{ab} V^a \ell^b]_H}.
\end{equation}
Furthermore the surface gravity is
\begin{equation}
\kappa_H = {1-2m'\over 2 r_H} = {1-8\pi(\rho_H+f_H)\over 2 r_H}
= {1-8\pi [T_{ab} V^a \ell^b]_H \over 2 r_H}.
\end{equation}
This now provides explicit formulae for the evolution and properties
of the horizon in terms of various quantities (densities, fluxes,
pressures, etc...) evaluated at the horizon.  The only real deficiency
in the tetrad we have chosen (and this is a matter of taste, not a
matter of physics), is that while it is easy to write down, and easy
to interpret, and easy to calculate with, the 4-velocity $V$ is not
geodesic (except in the rather special case where $c(r,\~t)\to c$ is a
constant). Now in that special case not only is $V$ geodesic, but also
$f=0$, the net flux seen by a freely falling observer located at the
evolving horizon is zero. This leads us to ask whether this behaviour
can be emulated by a slightly different choice of tetrad?

\section{General timelike-spacelike tetrad}
\label{S:generaltet}

Consider now the 4-velocity
\begin{equation}
\tilde V^a = {1\over c\sqrt{1-\beta^2}}\left[ 1, -(v+c\beta),0,0\right],
\end{equation}
where $|\beta(r,t)| < 1$. This corresponds to an observer who is
infalling with 3-velocity
\begin{equation}
v_3(r,t) \equiv v(r,t) + c(r,t) \; \beta(r,t).
\end{equation}
It is easy to check that $\tilde V$ is a unit timelike vector, and
easy to construct an appropriate spacelike vector that is orthonormal
to it:
\begin{equation}
\tilde S^a = {1\over c\sqrt{1-\beta^2}}\left[ -\beta, c+v\beta,0,0\right].
\end{equation}
This now allows us to construct a generic non-null tetrad (one element
of the tetrad, $\tilde V$, is guaranteed to be timelike, the other
three spacelike)
\begin{equation}
e_{\hat a}{}^a = 
\left( \tilde V^a, \tilde S^a, \hat\theta^a, \hat\phi^a \right).
\end{equation}
In terms of this tetrad one now has
\begin{eqnarray}
\fl\tilde\rho &= {G_{ab}\tilde V^a \tilde V^b\over8\pi}
\\
\fl & = {1\over4\pi r c (1-\beta^2)}
\left[
  c   (1-\beta^2) {m' \over r}
+ \left(\beta+ \sqrt{2m\over r} \right)^2 c'
+  \beta^2\partial_t\left(\sqrt{ 2m\over  r}\right)
\right];
\end{eqnarray}
\begin{eqnarray}
\fl
\tilde f &= {G_{ab}\tilde V^a \tilde S^b\over8\pi} \\
\fl
&= {1\over4\pi r c (1-\beta^2)} \;
\left[  -\beta\;\partial_t\left(\sqrt{ 2m\over  r}\right) -
\left\{ \beta+\sqrt{2m\over r} \right\} 
\left\{ 1+ \beta \sqrt{2m\over r} \right\} c'
\right];
\end{eqnarray}
\begin{eqnarray}
\fl
\tilde p_r &= {G_{ab}\tilde S^a \tilde S^b\over8\pi}
\\
\fl
&= {1\over4\pi r c (1-\beta^2)} \;
\left[
-c  (1-\beta^2) {m'\over r} + \left(1+\beta \sqrt{2m\over r} \right)^2 c'
+  \partial_t\left(\sqrt{ 2m\over  r}\right)
\right].
\end{eqnarray}
Then at the evolving horizon, where  $2m=r$, we have
\begin{eqnarray}
\tilde\rho_H &=  {1\over4\pi r c (1-\beta^2)}
\left[
c   (1-\beta^2) {m' \over r}
+ \left(1+ \beta \right)^2 c'
 +  \beta^2{\dot m\over r}
\right]_H;
\end{eqnarray}
\begin{eqnarray}
\tilde f_H &= {1\over4\pi r c (1-\beta^2)} \;
\left[  -\beta\;{\dot m \over r} -
\left( 1+ \beta\right)^2  c'
\right]_H;
\end{eqnarray}
\begin{eqnarray}
\tilde p_r^H
&= {1\over4\pi r c (1-\beta^2)} \;
\left[
-c   (1-\beta^2) {m'\over r} + \left(1+\beta \right)^2 c'
+ {\dot m \over r}
\right]_H.
\end{eqnarray}
The special case $\beta\to 0$ reproduces the results based on our
``natural'' tetrad.  From these results we can see that the notion of
``flux across the horizon'' is both subtle and less useful than one
might at first suppose --- the flux depends explicitly on the tetrad
(effectively, one's \emph{choice} of privileged observer), and indeed
the ``flux across the horizon'' can in many situations be made to
\emph{vanish} by choosing a suitable observer:
\begin{equation}
{\beta_H\over (1+\beta_H)^2} =  - \left[{r c'\over \dot m} \right]_H.
\end{equation}
But even though the notion of flux is somewhat more subtle than
expected, the evolution of the horizon as encoded in quantities such
as $\dot m|_H$ and $\d r_H/\d t$ is perfectly well defined and easy to
deal with.

\section{Formalism for generic horizons}
\label{S:generic}

We now show how to write the change in the area of the horizon as a
function of the expansions defined above. First, choose a coordinate
basis ($\theta, \phi$) for the surface of the horizon at a given
instant in time (on a given slicing). Then the area of this 2-surface
will be
\begin{equation}
{\cal A}_H = \int \sqrt{\det h}\;\d\theta \,\d\phi,
\end{equation}
where $h$ is the induced metric on the 2-sphere given by its
embedding in the full 4D spacetime.  For a spherically symmetric
2-sphere this is
\begin{equation} 
\d s^{2} = r^{2}(\d\theta^{2} + \sin^{2}\theta\;\d\phi^{2}),
\end{equation}
and thus $\sqrt{\det h} = r^{2}\sin\theta$ giving the
familiar $4\pi r^{2}$ area. Now if we ask how this area varies as
we move from one slice of the foliation to another we are led to
consider
\begin{equation}
\frac{{\mathrm{D}} {\cal A}_{H}}{\d\lambda} = 
\frac{\d x^{a}}{\d\lambda}\nabla_{a}{\cal A}_{H} = 
t^{a}\nabla_{a}{\cal A}_{H} =
{\cal{L}}_{t}{\cal A}_{H},
\end{equation}
where $\lambda$ is a parameter labelling the foliations, and $t^{a}$
is chosen to be tangential to the horizon, but normal to the foliation
and always future-pointing. To relate this expression to the
expansions of the ingoing and outgoing null rays rewrite the above as
\begin{equation} 4\pi {\cal{L}}_{t}r_{H}^{2} =
\frac{4\pi}{\sin\theta}\;
{\cal{L}}_{t}\sqrt{\det h} = 2\pi
r_{H}^{2}\; h^{ij}\;{\cal{L}}_{t}h_{ij}.
\end{equation}
Now identifying the intrinsic metric with the projection tensor
$q_{ab}$ and writing $q_{ab} = \hat \theta_{a}\,\hat \theta_{b} +
\hat\phi_{a}\,\hat\phi_{b}$ where $\hat\theta_{a}$ and $\hat\phi_{a}$
are two orthonormal directions tangent to the 2 surface, we get
\begin{equation}
\frac{{\mathrm{D}} {\cal A}_H}{\d\lambda} = 4\pi
r_{H}^{2} \; q^{ab}\nabla_{a}t_{b}. 
\end{equation}
Here
$t^{a}$ can always be written as a linear combination of
$\ell^{a}$ and $n^{a}$
\begin{equation} t^{a} = B\;\ell^{a} + C\;n^{a} \end{equation}
where $B$ and $C$ are the coefficients and depend in general on
$\lambda$. However on a given slice, for $t^{a}$ to be
future-pointing, there are only five cases to consider:\bigskip

\begin{tabular}{|c|c|c|c|c|}
\hline
    $t^a$ outside $\ell^{a}$ & $t^{a}\propto \ell^{a}$ & $t^{a}$ 
    between $\ell^{a}$ and $n^{a}$ & $t^{a}\propto n^{a}$ & $t^{a}$ 
    inside $n^{a}$ \\
  \hline
  $B>0$ & $B>0$ & $B>0$ & $B=0$ & $B<0$ \\
  $C<0$ & $C=0$ & $C>0$ & $C>0$ & $C>0$ \\
  \hline
\end{tabular}
\bigskip

\noindent Additionally, on the horizon,
\begin{equation} \frac{{\mathrm{D}} {\cal A}_H}{\d\lambda} = 4\pi
r_{H}^{2}C\theta_{n}.
\end{equation}
Therefore, since $\theta_{n}$ is negative at the horizon by
definition, the area ${\cal A}_H$ will be increasing in the first
case, constant in the second case, and decreasing in the other three
cases.

\section{Discussion}

We have demonstrated a simple framework capable of handling black
holes with rapidly evolving horizons.  The only physical simplifying
assumption was one of spherical symmetry, and the two mathematical
simplifying assumptions were the adoption of a spherically symmetric
slicing of spacetime, and the use of a specific nonsingular coordinate
chart (\Painleve--Gullstrand coordinates).  The framework does not
assume asymptotic flatness or stationarity, and we argue that these
concepts are not required to define a useful concept of horizon. While
exact spherical symmetry may not be an appropriate approximation to
describe astrophysical black holes, we feel that this framework is
useful in clarifying the basic concepts. In particular once someone
has understood the simple nature of the evolving horizon in
\Painleve--Gullstrand coordinates, it is difficult to then become
confused in other coordinate systems --- the ``frozen'' nature of the
horizon in Schwarzschild coordinates is then very clearly seen to be a
simple coordinate artefact due to that coordinate system becoming
degenerate on the horizon.

Furthermore, this framework is useful in that any investigation of the
black hole evaporation process \emph{must} ultimately consider the
essentially dynamic nature of the problem.  While a particular choice
of coordinates was used to facilitate some of the calculations, the
simplicity of the results should in no way be seen as an exclusive
property of the choice of coordinates.  We have made no attempt to
address the issue of singularity formation although Penrose's proof
will remain valid within its assumptions.  In this respect it is
important to note that the current work assumes a classical spacetime
everywhere, and one would expect it to arise as some sort of limiting
approximation to the models of Ashtekar and Bojowald who consider the
explicit breakdown of the spacetime manifold description.  To avoid a
singularity (as we certainly have trapped surfaces) we must either
violate the NEC (or have closed timelike curves, or not have a
manifold).  Conservative approaches to avoiding singularities focus on
the energy conditions and their violations. More radical approaches
(either within loop quantum gravity or string models) effectively
dispense with the spacetime manifold at sufficiently short distances.
Still, before adopting the more radical approaches it is useful to see
how far the standard manifold picture can be pushed.

One clear message that should be extracted form the current article is
that event horizons [at least in their precise mathematical definition
as absolute horizons] are \emph{not} essential for doing interesting
black hole physics.  Event horizons are mathematical abstractions that
in many ways encode and presuppose too many technical assumptions ---
assumptions which may or may not have anything to do with reality. For
physical black holes, evolving either due to accretion or Hawking
evaporation, other notions of horizon seem to be more appropriate.
Apparent horizons are sometimes useful but are also known to possess
several technical deficiencies. Isolated horizons are in many ways
``too close'' to being true event horizons (since no matter is allowed
to cross the isolated horizon it cannot either grow or shrink).
Hayward's trapping horizons, Ashtekar's dynamical horizons, and the
evolving horizons of this article, all capture key aspects of back
hole physics without presupposing the existence of an event horizon.
We would argue that the key feature of the evolving horizons of this
article is their relative simplicity and clarity, enabling one to
build up a simple physical picture and a clear intuition.

\appendix
\section{Coordinate basis}

For completeness, here are the coordinate basis components of the
Einstein tensor:
\begin{equation}
G_{tt} = {2m' c^2\over r^2} \left(1-{2m\over r}\right) 
+ {2\dot m c\over r^2 } {\sqrt{2m/r}};
\end{equation}
\begin{equation}
G_{tr} = {2\dot m\over r^2} - {2m' c\over r^2} \sqrt{2m\over r};
\end{equation}
\begin{equation}
G_{rr} = -{2m'\over r^2} +{2c'\over r c} 
+{2\dot m\over r^2 c} {1\over\sqrt{2m/r}};
\end{equation}
\begin{equation}
G_{\theta\theta} = G_{\hat\theta\hat\theta}\; r^2;
\end{equation}
\begin{equation}
G_{\phi\phi} = G_{\hat\theta\hat\theta}\; r^2 \; \sin^2\theta.
\end{equation}
At the evolving horizon
\begin{equation}
G_{tt} \to  {2\dot m c\over r^2 };
\end{equation}
\begin{equation}
G_{tr} \to {2(\dot m- m'c)\over r^2};
\end{equation}
\begin{equation}
G_{rr} \to {2c'\over r c} +{2(\dot m-m'c)\over r^2 c};
\end{equation}
and
\begin{equation}
\ell^a \to {(1,0,0,0)\over c}.
\end{equation}
This then easily yields (on the evolving horizon)
\begin{equation}
G_{ab}\ell^a\ell^b =   {2\dot m \over r^2 c},
\end{equation}
which is a simple consistency check on the entire formalism.

\section{Null decomposition}

Let the indices $i,j,\dots$ only take on the values $0,1$ (so they lie
in the $t$--$r$ plane). Then spherical symmetry is enough to imply
that
\begin{equation}
T_{ij} = T_0\; g_{ij} + T_+ \; \ell_i\;\ell_j + T_-\;n_i\;n_j.
\end{equation}
Then in terms of our natural tetrad
\begin{equation}
T_0 = {-\rho+p_r\over2}; \qquad T_+={\rho+p_r-2f\over4};\qquad  
T_-={\rho+p_r+2f\over4}.
\end{equation}
In terms of the Einstein tensor
\begin{equation}
G_0 = {2m'\over r} -{\dot m\over r c^2} {1\over\sqrt{2m/r}} 
- {c'\over r c} \left(1-{2m\over r}\right);
\end{equation}
\begin{equation}
G_+ = {\dot m\over2r^2 c} {1\over\sqrt{2m/r}} 
+ {c'\over2rc}\left(1+\sqrt{2m\over r}\right)^2;
\end{equation}
\begin{equation}
G_- = +{\dot m\over2r^2 c} {1\over\sqrt{2m/r}} 
+ {c'\over2rc}\left(1-\sqrt{2m\over r}\right)^2.
\end{equation}
At the apparent horizon
\begin{equation}
G_0 \to {2m'\over r} -{\dot m\over r c^2};
\end{equation}
\begin{equation}
G_+ \to {\dot m\over2r^2 c}  + {c'\over rc};
\end{equation}
\begin{equation}
G_- \to {\dot m\over2r^2 c}.
\end{equation}
If we now go to a general radially infalling tetrad $v_3 = v +
\beta c$, then $G_0$ is independent of $\beta$ whereas the null
vectors $\ell$ and $n$ are Doppler shifted
\begin{equation}
\ell \to \tilde \ell = \ell \; \sqrt{1-\beta\over1+\beta};
\qquad
n\to \tilde n = n \; \sqrt{1+\beta\over1-\beta};
\end{equation}
so that
\begin{equation}
G_+ \to G_+ \; {1+\beta\over1-\beta};
\qquad
G_- \to G_- \; {1-\beta\over1+\beta}.
\end{equation}
In particular the product $G_+\; G_-$ is invariant under choice of
radially infalling tetrad. This again demonstrates the while many
statements that one might wish to make concerning the evolving horizon
are tetrad dependent, there are also a number of useful quantities
that are tetrad-independent. In particular, note that if the evolving
horizon is momentarily static, then since $\dot m = 0$ we have $G_-=0$,
and the stress-energy on the horizon exhibits a type of ``enhanced
symmetry''~\cite{enhanced} in that it takes on the form
 \begin{equation}
\label{E:reduced}
T_{ij} = T_0\; g_{ij} + T_+ \; \ell_i\;\ell_j.
\end{equation}
Under the significantly stronger assumption that the evolving horizon
is static, at least for some finite time interval, and has been static
long enough for a bifurcation 2-surface to form, then since $\ell\to0$
on the bifurcation 2-surface, we have the much stronger result that
\begin{equation}
\label{E:enhanced}
T_{ij} \to T_0\; g_{ij}
\end{equation}
on the bifurcation 2-surface. This makes it clear that the ``enhanced
symmetries'' of reference~\cite{enhanced} are very much related to the
existence of both a finite-time translation symmetry and a bifurcation
2-surface. The enhanced symmetry (\ref{E:enhanced}) is already reduced
as one moves away from the bifurcation 2-surface, and even the reduced
symmetry (\ref{E:reduced}) is at best only approximate once the
horizon starts evolving.

\section{Special case: $c(r,t)=1$}

There is a special case of the formalism in which technical
computations simplify even further --- that is if we set $c(r,t)=1$.
This is not a situation motivated by any particular physical
principle: Although it is true that $c(r,t)=1$ for both the
Schwarzschild and Reissner--{\Nordstrom} solutions, there does not
appear to be any deep reason for this. (The analogous property also
holds for the Doran form of the Kerr solution~\cite{Doran}, though
again there seems to be no deep physical reason underlying this.)
Nevertheless, the tremendous simplifications attendant on setting
$c(r,t)=1$ are so significant, and do not seem to undermine the main
points we are making regarding the non-perturbative evolution of
horizons, that a brief summary may be warranted. (See the analogous
discussion by Hayward in~\cite{Hayward-special}.)

Suppose we consider the metric:
\begin{equation} \label{PGmetric2}
\d s^2 = - [1-v(r,\~t)^2]\d \~t^2 + 2 v(r,\~t) \; \d r\; \d \~t 
+ \d r^2 + r^2 \; \d\Omega^2,
\end{equation}
and set $v(r)=\sqrt{2m(r,\~t)/r}$. Then the discussion of the evolving
horizon is largely unaltered though now we have the simplification
\begin{equation}
\kappa_\ell = \kappa_n = - v'(r,t) = {m/r^2-m'\over\sqrt{2m/r}}.
\end{equation}
In fact, we now have
\begin{equation}
G_{ab} \ell^a \ell^b = G_{ab} n^a n^b,
\end{equation}
which is enough to tell us without detailed computation that
\begin{equation}
G_{ab} V^a S^b = 0.
\end{equation}
Indeed the ``comoving'' tetrad defined by $V^a$ and $S^a$ is now
geodesic, and the stress energy tensor in this tetrad basis simplifies
to:
\begin{equation}
\rho = {G_{\hat t\hat t}\over8\pi} = {m'\over 4\pi r^2};
\end{equation}
\begin{equation}
f = {G_{\hat t\hat r}\over8\pi} = 0;
\end{equation}
\begin{equation}
p_r = {G_{\hat r\hat r}\over8\pi}  = -{m'\over 4\pi r^2}  
+ { \dot m\over 4\pi  r^2} {1\over\sqrt{2m/r}}.
\end{equation}
The only ``complicated'' component of the Einstein tensor is the
transverse one:
\begin{eqnarray}
\fl
p_t = {G_{\hat\theta\hat\theta}\over8\pi} 
= {G_{\hat\phi\hat\phi}\over8\pi} &=&
-{m''\over 8\pi r} +\sqrt{2m\over r} \left[ {\dot m\over32 \pi m^2r} 
\left[ m-rm'\right] + {1\over16m\pi} \dot m'  \right].
\end{eqnarray}
At the apparent horizon  the only real simplifications are
\begin{equation}
p_{r,H} =
-\left[{m' - \dot m \over 4\pi r^2}  \right]_H,
\end{equation}
and
\begin{eqnarray}
p_{t,H} &=&
\left[{ - m m'' + m \dot m' + \dot m(1-2m')/4\over 4\pi r^2}\right]_H.
\end{eqnarray}
This particular class of spacetime geometries is slightly easier to
deal with than those considered in the bulk of the article, and so is
perhaps of some mathematical and pedagogical interest.  However it
should be noted that the more fundamental issues of interest in this
article are the way in which we have explicitly shown how a suitable
choice of coordinates completely side-steps the ``frozen'' nature of
coordinate time at the Schwarzschild horizon, and allows one to
non-perturbatively formulate questions about the production and decay
of evolving horizons.

\section{Vaidya Spacetime}

The standard expression of the accreting Vaidya spacetime \cite{Vaidya,
  Exact} in advanced null coordinates is
\begin{equation} 
\d s^{2} = -\left(1-\frac{2m(v)}{r}\right)\d v^{2} +2\,\d v\,\d r
+ r^{2}\,\d\Omega^{2}.
\end{equation}
\setcounter{footnote}{0} This spacetime represents collapsing null
radiation, and the function $m(v)$ is an arbitrary non-decresing function
of its argument.\;\footnote{There is a related time-reversed ``shining
  star'' solution for outgoing null radiation, which can be written as
\[
\d s^{2} = -\left(1-\frac{2m(u)}{r}\right)\d u^{2} -2\,\d u\,\d r
+ r^{2}\,\d\Omega^{2},
\]
where $u$ is a retarded null coordinate and $m(u)$ is a non-increasing
function of its argument. This can also be used as a model for a
rather specific class of evaoprating black hole. The comments below
can immediately immediately be carried over to this case by suitable
changes in notation. } Note that the mass function is a function
\emph{only} of the advanced null coordinate
$v$.\;\footnote{Unfortunately the null coordinate $v$ occurring here
  has nothing to do with the function $v(r,\tilde t)$ previously used
  to discuss the \Painleve--Gullstrand form of the metric.  The usage
  and notation is unfortunately standard, and care must be taken to
  keep the two concepts distinct.}  We can transform to
\Painleve--Gullstrand coordinates ($g_{rr}=1$) by writing the null
coordinate $v$ as a function of \Painleve--Gullstrand coordinates
$\tilde t$ and $r$: that is, write $v = v(\tilde t, r)$, whence
\begin{equation} 
\d v = \frac{\partial v}{\partial \tilde t}\;\d \tilde t 
+ \frac{\partial v}{\partial r} \;\d r.
\end{equation}
Substituting, we obtain
\begin{eqnarray}
\label{vai1} \d s^{2} & = &
-\left(1-\frac{2m(v(\tilde t,r))}{r}\right)\left(
\frac{\partial v}{\partial \tilde t}
\right)^{2}\d \tilde t^{2} 
\nonumber \\ & &  +
\left[
-2\left(1-\frac{2m(v(\tilde t,r))}{r}\right)
\left(\frac{\partial v}{\partial \tilde t}\right)
\left(\frac{\partial v}{\partial r}\right) +
2\left(\frac{\partial v}{\partial \tilde t}\right)\right]
\d \tilde t\,\d r
\nonumber \\ & & +
\left[-\left(1-\frac{2m(v(\tilde t,r))}{r}\right)
\left(\frac{\partial v}{\partial r}\right)^{2} 
+ 2\left(\frac{\partial v}{\partial r}\right)\right]\d r^{2} 
+ r^{2}\d\Omega^{2}. 
\end{eqnarray}
To obtain \Painleve--Gullstrand coordinates we demand $g_{rr}=1$ so
\begin{equation}
\left[-\left(1-\frac{2m(v(\tilde t,r))}{r}\right)
\left(\frac{\partial v}{\partial r}\right)^{2} 
+ 2\left(\frac{\partial v}{\partial
r}\right)\right] = 1.
\end{equation}
Solving this quadratic yields a first-order differential equation
relating the coordinates $(v,r)$ and $(\tilde t, r)$:
\begin{equation} 
\label{E:ode}
\frac{\partial v}{\partial r} = \frac{1 \pm
\sqrt{\frac{2m(v(\tilde t,r))}{r}}}{1-\frac{2m(v(\tilde t,r))}{r}}
= {1\over 1 \mp \sqrt{\frac{2m(v(\tilde t,r))}{r}} }.
\end{equation}
Since $v$ is by construction regular at the evolving horizon we must
take $\pm\to-$ and $\mp\to+$ above.  Putting this back in to equation
(\ref{vai1}) yields the \Painleve--Gullstrand form of the Vaidya
solution
\begin{eqnarray}
 \d s^{2} 
& = &
-\left(1-\frac{2m(v(\tilde t,r))}{r}\right)
\left(\frac{\partial v}{\partial \tilde t}\right)^{2}\d\tilde  t^{2} 
\nonumber \\ & & 
+ 2
\sqrt{\frac{2m(v(\tilde t,r))}{r}}
\left(\frac{\partial v}{\partial \tilde t}\right)
\,\d \tilde t\,\d r 
+ \d r^{2} + r^{2}\d\Omega^{2}. 
\end{eqnarray}
In the original null coordinates the evolving horizon occurs at 
\begin{equation}
r_H(v) = 2 m(v),
\end{equation}
and there is clearly only a single horizon for any value of the null
coordinate $v$. In terms of the \Painleve--Gullstrand coordinates
$(\tilde t, r)$ the location of the horizon is given implicitly by
\begin{equation}
r_H(v(\tilde t,r)) = 2 m(v(\tilde t, r)),
\end{equation}
where $v(\tilde t, r)$ is constrained by the differential equation
\eref{E:ode}. If the horizon is spacelike this condition can in
principle have many solutions for fixed $\tilde t$, as the horizon
can in principle move back and forth in \Painleve--Gullstrand time.
Consider in particular the quantity ${\partial m(v(\tilde t,
  r))}/{\partial r}$:
\begin{equation} 
\frac{\partial m}{\partial r} = \frac{\d m}{\d
v}\frac{\partial v}{\partial r} = \frac{\d m}{\d v} 
{1 \over 1 + \sqrt{\frac{2m(t,r)}{r}}}. 
\end{equation}
Then as $2m \rightarrow r$ this becomes
\begin{equation} 
\frac{\partial m}{\partial r} \to m'_H = 
\frac{1}{2}\frac{\d m}{\d v}. 
\end{equation}
In order to violate equation (\ref{extremeBH}) we would need $m'_H >
1/2$.  But since in the Vaidya solution ${\d m}/{\d v}$ can be
\emph{any} non-negative function, this condition can certainly be
violated.  In short, for a spacelike evolving horizon ``outermost'' in
the sense of the $(v,r)$ null coordinates may not always coincide with
``outermost'' in the sense of the $(\tilde t, r)$
\Painleve--Gullstrand coordinates. We emphasise that this is not an
inconsistency in the formalism, merely one of the interesting
coordinate artefacts one has to keep in mind.

\section*{Acknowledgements}

This research was supported by the Marsden Fund administered by the
Royal Society of New Zealand. We wish to thank the referees for useful
comments and suggestions.

\section*{References}



\end{document}